\documentclass[superscriptaddress,prc,aps,twocolumn]{revtex4}
\usepackage{amsmath,amssymb}
\usepackage{graphicx}
\usepackage{hyperref}
\newcommand{\Mainz}[1]
{\affiliation{Institut f\"ur Kernphysik, University of Mainz, D-55099 Mainz,Germany}}
\newcommand{\Bonn}[1]
{\affiliation{Helmholtz-Institut f\"ur Strahlen- und Kernphysik, University of Bonn,
 D-53115 Bonn, Germany}}
\newcommand{\Regina}[1]
{\affiliation{University of Regina, Regina, Saskatchewan S4S 0A2, Canada}}
\newcommand{\Kent}[1]
{\affiliation{Kent State University, Kent, Ohio 44242-0001, USA}}
\newcommand{\Glasgow}[1]
{\affiliation{SUPA School of Physics and Astronomy, University of Glasgow,
 Glasgow G12 8QQ, United Kingdom}}
\newcommand{\Dubna}[1]
{\affiliation{Joint Institute for Nuclear Research, 141980 Dubna, Russia}}
\newcommand{\Pavia}[1]
{\affiliation{INFN Sezione di Pavia, I-27100 Pavia, Italy}}
\newcommand{\GWU}[1]
{\affiliation{The George Washington University, Washington, DC 20052-0001, USA}}
\newcommand{\Basel}[1]
{\affiliation{Institut f\"ur Physik, University of Basel, CH-4056 Basel, Switzerland}}
\newcommand{\York}[1]
{\affiliation{Department of Physics, University of York, Heslington, York, Y010 5DD, United Kingdom}}
\newcommand{\INR}[1]
{\affiliation{Institute for Nuclear Research, 125047 Moscow, Russia}}
\newcommand{\Sackville}[1]
{\affiliation{Mount Allison University, Sackville, New Brunswick E4L 1E6, Canada}}
\newcommand{\Zagreb}[1]
{\affiliation{Rudjer Boskovic Institute, HR-10000 Zagreb, Croatia}}
\newcommand{\Amherst}[1]
{\affiliation{University of Massachusetts, Amherst, Massachusetts 01003, USA}}
\begin{document}
\title{
New high-statistics measurement of the $\pi^0 \to e^+e^-\gamma$ Dalitz decay at the Mainz Microtron}

\author{S.~Prakhov}\thanks{Contact author: prakhov@ucla.edu}\Mainz \\
\author{L.~Heijkenskj\"old}\Mainz \\
\author{S.~Abt}\Basel \\ 
\author{P.~Achenbach}\Mainz \\
\author{P.~Adlarson}\Mainz \\
\author{F.~Afzal}\Bonn \\
\author{Z.~Ahmed}\Regina \\
\author{K.~Altangerel}\Mainz \\
\author{J.~R.~M.~Annand}\Glasgow \\
\author{M.~Bashkanov}\York \\
\author{R.~Beck}\Bonn \\
\author{M.~Biroth}\Mainz \\
\author{N.~S.~Borisov}\thanks{Deceased}\Dubna \\
\author{A.~Braghieri}\Pavia \\
\author{W.~J.~Briscoe}\GWU \\
\author{F.~Cividini}\Mainz \\
\author{C.~Collicott}\Mainz \\
\author{S.~Costanza}\Pavia \\
\author{A.~Denig}\Mainz \\
\author{M.~Dieterle}\Basel \\
\author{A.~S.~Dolzhikov}\Dubna \\
\author{E.~J.~Downie}\GWU \\
\author{P.~Drexler}\Mainz \\
\author{S.~Fegan}\York \\
\author{S.~Gardner}\Glasgow \\
\author{D.~Ghosal}\Basel \\
\author{D.~I.~Glazier}\Glasgow \\
\author{I.~Gorodnov}\Dubna \\
\author{W.~Gradl}\Mainz \\
\author{M.~G\"unther}\Basel \\
\author{G.~M.~Gurevich}\INR \\
\author{D.~Hornidge}\Sackville \\
\author{G.~M.~Huber}\Regina \\
\author{A.~K\"aser}\Basel\\
\author{V.~L.~Kashevarov}\Mainz \\ \Dubna \\
\author{S.~J.~D.~Kay}\Regina \\
\author{M.~Korolija}\Zagreb \\
\author{B.~Krusche}\thanks{Deceased}\Basel \\
\author{A.~Lazarev}\Dubna \\
\author{K.~Livingston}\Glasgow \\
\author{S.~Lutterer}\Basel \\
\author{I.~J.~D.~MacGregor}\Glasgow \\
\author{D.~M.~Manley}\Kent \\ 
\author{P.~P.~Martel}\Mainz \\
\author{R.~Miskimen}\Amherst \\
\author{M.~Mocanu}\York \\
\author{E.~Mornacchi}\Mainz \\
\author{C.~Mullen}\Glasgow \\
\author{A.~Neganov}\Dubna \\
\author{A.~Neiser}\Mainz \\ 
\author{M.~Ostrick}\Mainz \\  
\author{P.~B.~Otte}\Mainz \\
\author{D.~Paudyal}\Regina \\
\author{P.~Pedroni}\Pavia \\
\author{A.~Powell}\Glasgow \\
\author{E.~Rickert}\Mainz \\
\author{T.~Rostomyan}\Basel \\
\author{V.~Sokhoyan}\Mainz \\
\author{K.~Spieker}\Bonn \\
\author{O.~Steffen}\Mainz \\
\author{I.~I.~Strakovsky}\GWU \\
\author{Th.~Strub}\Basel \\
\author{I.~Supek}\Zagreb \\
\author{M.~Thiel}\Mainz \\
\author{A.~Thomas}\Mainz \\   
\author{Yu.~A.~Usov}\Dubna \\
\author{S.~Wagner}\Mainz \\
\author{D.~P.~Watts}\York \\
\author{D.~Werthm\"uller}\York \\
\author{J.~Wettig}\Mainz \\
\author{M.~Wolfes}\Mainz \\
\author{N.~Zachariou}\York \\

\collaboration{A2 Collaboration at MAMI}

\date{\today}
         
\begin{abstract}
 The Dalitz decay $\pi^0 \to e^+e^-\gamma$ has
 been measured with the highest statistical accuracy obtained so far in
 the $\gamma p\to \pi^0 p$ reaction with the A2 tagged-photon facility at the Mainz Microtron, MAMI.
 The value of the slope parameter for the $\pi^0$ electromagnetic transition form factor,
 $a_\pi=0.0315\pm 0.0026_{\mathrm{stat}}\pm 0.0010_{\mathrm{syst}}$,
 is obtained from the analysis of $2.4\times10^6$ $\pi^0 \to e^+e^-\gamma$ observed decays.
 Within experimental uncertainties, it is in agreement with existing measurements and
 theoretical calculations, with its own uncertainty being smaller than previous
 results based on the analysis of $\pi^0\to e^+e^-\gamma$ decays.
\end{abstract} 

\maketitle

\section{Introduction}

 The importance of measuring electromagnetic (e/m) transition form factors (TFFs)
 of light mesons, and especially $\pi^0$, for better understanding their properties
 and providing low-energy precision tests of the Standard Model (SM)
 and Quantum Chromodynamics (QCD), has been discussed in some detail
 in Ref.~\cite{Pi0_TFF_A2_2016} dedicated to the previous A2 measurement
 of the $\pi^0 \to e^+e^-\gamma$ Dalitz decay.
 In particular, for data-driven theoretical determinations of the anomalous
 magnetic moment of the muon $(g-2)_\mu$ within the SM~\cite{Nyffeler_2016,g_2,g_2_u},
 the TFFs of light mesons enter as important contributions to the hadronic
 light-by-light (HLbL) scattering calculations~\cite{Colangelo_2014,Colangelo_2015,Masjuan_2017,Colangelo_2017,Hoferichter_2018,Hoferichter_2018_2,Eichmann_2019,Bijnens_2019,Leutgeb_2020,Cappiello_2020,Masjuan_2022,Bijnens_2020,Bijnens_2021,Danilkin_2012,Stamen_2022,Leutgeb_2023,Hoferichter_2023,Hoferichter_2024,Estrada_2024,Ludtke_2025,Deineka_2025,Eichmann_2025,Bijnens_2025,Hoferichter_2025,Holz_2025,Cappiello_2025}.
 If the TFF parameters are extracted from the $\pi^0 \to e^+e^-\gamma$ Dalitz decay
 with a sufficient accuracy, they could then
 constrain calculations that estimate the pion-pole term $a_{\mu}^{\pi^0}$
 to the HLbL scattering contribution to $(g-2)_\mu$, as well as the $\pi^0\gamma$-channel
 contribution $a_{\mu}^{\pi^0\gamma}$~\cite{Jegerlehner_2017,Davier_2017,Keshavarzi_2018,Keshavarzi_2020,Davier_2020,Hoid_2020}
 to the hadronic-vacuum-polarization (HVP)
 correction~\cite{Bouchiat_1961,Brodsky_1968} to $(g-2)_\mu$.
    
 For a structureless (point-like) meson $A$, its decays into a lepton pair plus
 a photon, $A\to l^+l^-\gamma$, can be described within
 Quantum Electrodynamics (QED) via $A\to \gamma^*\gamma$, with the virtual photon $\gamma^*$
 decaying into the lepton pair~\cite{QED}. For the meson $A$, QED predicts a specific strong
 dependence of its decay rate on the dilepton invariant mass, $m^2_{ll}=q^2$.
 A deviation from the pure QED dependence, caused by the actual electromagnetic
 structure of the meson $A$, is formally described by its e/m TFF~\cite{Landsberg}.
 The Vector-Meson-Dominance (VMD) model~\cite{Sakurai} can be used to describe the coupling of
 the virtual photon $\gamma^*$ to the meson $A$ via an intermediate virtual vector meson $V$.
 This mechanism is especially strong in the time-like (energy transfer larger than
 the momentum transfer) momentum-transfer region, $(2m_l)^2 < q^2 < m_A^2$, where
 a resonant behavior near $q^2 = m^2_V$ of the virtual photon arises because
 the virtual vector meson is approaching the mass shell~\cite{Landsberg}.
 Experimentally, time-like TFFs can be determined via measuring the differential decay width
 of $A\to l^+l^-\gamma$ as a function of the dilepton invariant mass $m_{ll}$,
 normalizing this dependence to the partial decay width $\Gamma(A\to \gamma\gamma)$,
 and taking into account the pure QED dependence expected for
 the $A\to \gamma^*\gamma \to l^+l^-\gamma$ differential decay width.
 Correspondingly, the $\pi^0\to \gamma^*\gamma \to e^+e^-\gamma$ differential decay width can
 be parametrized in the time-like region as~\cite{Landsberg}
\begin{eqnarray}
 & & \frac{d\Gamma(\pi^0\to e^+e^-\gamma)}{dm_{ee}\Gamma(\pi^0\to \gamma\gamma)} =
 [{\rm{QED}}(m_{ee})] |F_{\pi^0\gamma}(m_{ee})|^2
\nonumber 
\\ 
 & = & \frac{4\alpha}{3\pi m_{ee}} \times \bigg(1-\frac{4m^2_e}{m^2_{ee}}\bigg)^{\frac{1}{2}}
\bigg(1+\frac{2m^2_e}{m^2_{ee}}\bigg) \bigg(1-\frac{m^2_{ee}}{m^2_{\pi^0}}\bigg)^{3}
\nonumber 
\\ 
 & \times & |F_{\pi^0\gamma}(m_{ee})|^2 ,
\label{eqn:dgdm_pi0}
\end{eqnarray}
 where $F_{\pi^0\gamma}$ is the normalized TFF of the $\pi^0$ meson, $m_{\pi^0}$ and
 $m_{e}$ are the masses of the $\pi^0$ meson and $e^{\pm}$, respectively.
 Because of the smallness of the momentum-transfer range for the $\pi^0\to e^+e^-\gamma$
 decay, its normalized TFF is typically parametrized as~\cite{PDG_2024}
\begin{equation}
 F_{\pi^0\gamma}(m_{ee}) = 1 + a_\pi\frac{m^2_{ee}}{m^2_{\pi^0}},
\label{eqn:Fm}
\end{equation}
 where the parameter $a_\pi$ reflects the TFF slope at $m^2_{ee}=0$.
 A simple VMD model incorporates only the $\rho$, $\omega$, and $\phi$
 resonances (in the narrow-width approximation) as virtual vector mesons
 driving the photon interaction in $A\to\gamma^*\gamma$.
 Using a quark model for the corresponding couplings leads to neglecting $\phi$
 and yields~\cite{Landsberg}
 $a_\pi/m^2_{\pi^0} = 0.5 (1 + m^2_{\rho}/m^2_{\omega})/m^2_{\rho}\approx 1.648$~GeV$^{-2}$ 
 (or $a_\pi \approx 0.0300$) for the $\pi^0$ Dalitz decay.

 Another feature of the $\pi^0\to e^+e^-\gamma$ decay amplitude is an angular anisotropy
 of the virtual photon decaying into the $e^+e^-$ pair.
 For the $e^+$, $e^-$, and $\gamma$ in the $\pi^0$ rest frame,
 the angle $\theta_e$ between the direction of one of the leptons
 in the virtual-photon (or dilepton) rest frame and the direction
 of the dilepton system (which is opposite to the $\gamma$ direction)
 follows the dependence~\cite{NA60_2016}
\begin{equation}
 f(\cos\theta_e) = 1 + \cos^2\theta_e + \bigg(\frac{2m_e}{m_{ee}}\bigg)^2 \sin^2\theta_e
\label{eqn:dtheta}
\end{equation}
 with the $\sin^2\theta_e$ term becoming very small when $m_{ee}\gg 2m_e$.
 Both the $[{\rm{QED}}(m_{ee})]$ term in Eq.(\ref{eqn:dgdm_pi0}) and
 the angular dependence in Eq.(\ref{eqn:dtheta}) represent only
 the leading-order term of the $\pi^0\to e^+e^-\gamma$ decay amplitude,
 and radiative corrections need to be considered for a more accurate
 calculation of $[{\rm{QED}}(m_{ee},\cos\theta_e)]$. The introduction
 of radiative corrections modifies Eq.(\ref{eqn:dgdm_pi0}) as
\begin{equation}
 \frac{d\Gamma(\pi^0\to e^+e^-\gamma)}{dm_{ee}\Gamma(\pi^0\to \gamma\gamma)} =
 [{\rm{QED}}(m_{ee})] (1+\delta(m_{ee})) |F_{\pi^0\gamma}(m_{ee})|^2,
\label{eqn:dgdm_pi0_rad}
\end{equation}
 where $\delta(m_{ee})$ is radiative correction as a function of $m_{ee}$
 intergated over the full $\cos\theta_e$ range.
 The most recent calculations of radiative corrections to the differential
 decay rate of the Dalitz decay $\pi^0\to e^+e^-\gamma$ were reported
 in Ref.~\cite{Husek_2015}. In this work, the radiative corrections derived
 by Mikaelian and Smith~\cite{MS_1972} to the leading-order (LO) differential decay
 rate of the Dalitz decay $\pi^0\to e^+e^-\gamma$ were recalculated beyond
 the soft-photon approximation. The next-to-leading-order (NLO) corrections
 were divided into three parts emphasizing their origin. They included
 the virtual radiative corrections coming from both the electron and muon loops,
 the one-photon-irreducible contribution at one-loop level,
 and the bremsstrahlung correction.

 The first high-statistics measurements of the time-like $\pi^0$ TFF were conducted
 somewhat recently by the A2~\cite{Pi0_TFF_A2_2016} and NA62~\cite{Pi0_TFF_NA62_2017}
 Collaborations via the analysis of $0.4\times10^6$ and $1.1\times10^6$ Dalitz decays
 $\pi^0\to e^+e^-\gamma$, respectively. Before that, 
 the magnitude of the Dalitz-decay slope parameter $a_\pi$ and its uncertainty
 in the Review of Particle Physics (RPP) from 2014~\cite{PDG_2014}, $a_\pi = 0.032\pm0.004$,
 were mostly determined by a measurement of the space-like $\pi^0$ TFF in
 the process $e^+e^-\to e^+e^-\pi^0$ by the CELLO detector~\cite{CELLO_1991}.
 Adding the A2 value $a_\pi = 0.030\pm 0.010_{\mathrm{tot}}$~\cite{Pi0_TFF_A2_2016} and
 $a_\pi = 0.0368\pm 0.0051_{\mathrm{stat}}\pm 0.0025_{\mathrm{syst}} = 0.0368\pm 0.0057_{\mathrm{tot}}$
 from the NA62~\cite{Pi0_TFF_NA62_2017} measurement~\cite{Pi0_TFF_NA62_2017}
 in the RPP average resulted in its new value $a_\pi = 0.0332\pm0.0029$~\cite{PDG_2024},
 which is slightly larger then the previous RPP value, mostly due to smaller uncertainties
 in the NA62 result.

 Quite recent theoretical calculations for the $\pi^0\to \gamma^*\gamma \to e^+e^-\gamma$ TFF,
 in addition to the slope parameter $a_\pi$, also involve the curvature
 parameter $b_\pi$:
\begin{equation}
 F_{\pi^0\gamma}(m_{ee}) = 1 + a_\pi\frac{m^2_{ee}}{m^2_{\pi^0}}+ b_\pi\frac{m^4_{ee}}{m^4_{\pi^0}}~.
\label{eqn:Fm2}
\end{equation}
 A calculation based on a model-independent method using Pad\'e approximants
 was reported in Ref.~\cite{Mas12}.
 The analysis of space-like data (CELLO~\cite{CELLO_1991}, CLEO~\cite{CLEO_1998},
 BABAR~\cite{BABAR_2011}, and Belle~\cite{Belle_2012}) with this method provides
 a good and systematic description of the low-energy region, resulting in
 $a_\pi = 0.0324\pm0.0012_{\mathrm{stat}}\pm0.0019_{\mathrm{syst}}$ and
 $b_\pi = (1.06\pm0.09_{\mathrm{stat}}\pm0.25_{\mathrm{syst}})\times 10^{-3}$.
 Adding the latest time-like data in this method will definitely constrain these calculations
 to a better accuracy.
 Values with even smaller uncertainties, $a_\pi = 0.0307\pm0.0006$ and
 $b_\pi = (1.10\pm0.02)\times 10^{-3}$, were obtained by using
 dispersion theory~\cite{Hoferichter_2014}.
 In that analysis, the singly virtual TFF was calculated in both
 the time-like and the space-like regions, based
 on data for the $e^+e^-\to 3\pi$ cross section, generalizing previous
 studies on $\omega/\phi \to 3\pi$ decays~\cite{Niecknig_2012} and
 $\gamma\pi \to \pi\pi$ scattering~\cite{Hoferichter_2012},
 and verifying the results by comparing them to time-like $e^+e^-\to \pi^0\gamma$ data
 at larger momentum transfer. This analysis was later revised to match better
 the high-energy asymptotics that become more relevant once the TFF is incorporated
 into the HLbL loop integral for the $(g-2)_\mu$ contribution, which resulted
 in the new value $a_\pi = 0.0315\pm0.0009$~\cite{Hoferichter_2018,Hoferichter_2018_2}.
  
 In this paper, a new high-statistics measurement of the $\pi^0\to e^+e^-\gamma$ Dalitz
 decay with the A2 experimental setup at MAMI is reported.    
 The experiment was conducted in 2018 by producing $\approx 3.31\times10^9$ $\pi^0$ mesons
 in the reaction $\gamma p\to \pi^0 p$, which made it possible to accumulate
 $\approx2.3\times10^6$ $\pi^0\to e^+e^-\gamma$ decays available for the experimental analysis,
 compared to $\approx0.4\times 10^6$ decays used in the previous measurement
 by the A2 Collaboration~\cite{Pi0_TFF_A2_2016}.
 The $\pi^0$ TFF parameters and their uncertainties extracted in the present measurement
 represent a further and more accurate experiment in the time-like momentum-transfer region
 that aims to better constrain calculations estimating $(g-2)_\mu$ contributions
 from the pion-pole term $a_{\mu}^{\pi^0}$ to the HLbL scattering and the $\pi^0\gamma$
 term $a_{\mu}^{\pi^0\gamma}$ to the HVP correction.
  
\section{Experimental setup}
\label{sec:Setup}

 The process $\gamma p\to \pi^0 p \to e^+e^-\gamma p$ was measured
 at the the Mainz Microtron (MAMI)~\cite{MAMI,MAMIC}, using an energy-tagged bremsstrahlung
 photon beam. The energies of the incident photons were analyzed up to 750~MeV,
 by detecting the post-bremsstrahlung electrons in the Glasgow--Mainz tagged-photon spectrometer
 (Tagger)~\cite{Tagger_Old}, detectors and electronics of which underwent a major upgrade
 in 2017~\cite{Tagger_New}.
 The uncertainty of $\pm 1.5$~MeV in the energy of the tagged photons was mostly determined by
 the segmentation of the focal-plane detector of the Tagger in combination with the energy
 of the MAMI electron beam, which was set for the present experiment at 883~MeV.

 The final-state particles produced in the $\gamma p$ interactions were detected by using
 the Crystal Ball (CB)~\cite{CB} as a central calorimeter
 and TAPS~\cite{TAPS,TAPS2} as a forward calorimeter
 that was installed 1.5~m downstream of the CB center.  
 The CB detector consists of 672 NaI(Tl) crystals covering polar angles from $20^{\circ}$
 to $150^{\circ}$. The TAPS calorimeter consists of 366 BaF$_2$ crystals covering polar angles
 from $4^{\circ}$ to $20^{\circ}$ and 72 PbWO$_{4}$ crystals with angular coverage from
 $1^{\circ}$ to $4^{\circ}$. Both the CB and TAPS calorimeters have full azimuthal coverage.
 More information regarding the energy and angular resolution of the CB and TAPS detectors
 is provided in Refs.~\cite{etamamic,slopemamic}.

 Linear-polarized photons, produced by coherent bremsstrahlung of the MAMI beam electrons
 in a 100-$\mu$m-thick diamond radiator and collimated by a 3-mm-diameter Pb collimator,
 were incident on a 10-cm-long liquid hydrogen (LH$_2$) target located
 in the center of the CB. The coherent peak was positioned at the center-of-mass
 energy corresponding to the maximum of the $\Delta(1232)$ resonance, enhancing
 significantly pion photoproduction. The polarization orientation was not relevant
 for the present measurement.
 The total amount of material around the LH$_2$ target,
 including the Kapton cell and the 1-mm-thick carbon-fiber beamline,
 was equivalent to 0.8\% of a radiation length $X_0$.
 In the present measurement, it was essential to keep the material budget
 as low as possible to minimize the background from $\pi^0 \to \gamma\gamma$ decays
 with conversion of the photons into $e^+e^-$ pairs.
 The LH$_2$ target was surrounded by a Particle IDentification
 (PID) detector~\cite{PID} used to distinguish between charged and
 neutral particles. It was made of 24 scintillator bars
 (50 cm long, 4 mm thick) arranged as a cylinder with a radius of 12 cm.
 The experimental trigger required the total energy deposited in the CB
 to exceed $\approx$140~MeV.
\begin{figure}
\includegraphics[width=0.5\textwidth]{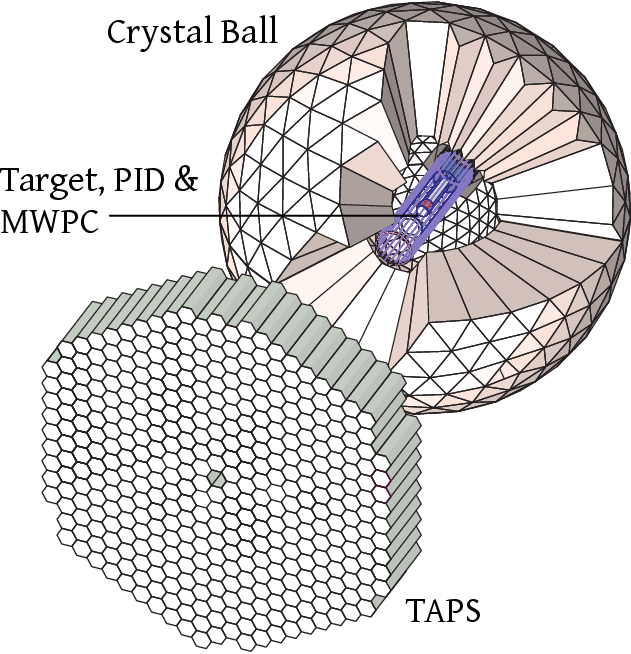}
\caption{(Color online)
  A general sketch of the Crystal Ball, TAPS, and PID detectors.
}
 \label{fig:cb_taps_pid} 
\end{figure}

 A general sketch of the CB, TAPS, and PID is shown
 in Fig.~\ref{fig:cb_taps_pid}.
 A multi-wire proportional chamber, MWPC, also shown in this figure
 (which consists of two cylindrical MWPCs inside each other),
 was not used in the present measurements because of its relatively low
 efficiency for detecting $e^\pm$.

 \section{Data handling}
\label{sec:Data}
\subsection{Event selection}
\label{subsec:Data-I}

 To search for a signal from $\pi^0 \to e^+e^-\gamma$ decays, 
 candidates for the process $\gamma p\to e^+e^-\gamma p$
 were extracted from events having three or four clusters
 reconstructed by a software analysis in the CB and TAPS together.
 The procedure used in the present analysis for the event selection
 was quite similar to the previous A2 measurement of this decay reported
 in Ref.~\cite{Pi0_TFF_A2_2016}.
 The offline cluster algorithm was optimized for finding
 a group of adjacent crystals in which the energy was deposited
 by a single-photon e/m shower. This algorithm works well for $e^{\pm}$,
 which also produce e/m showers in the CB and TAPS, and for proton clusters.
 The software threshold for the cluster energy was chosen to be 12 MeV.
 For the $\gamma p\to e^+e^-\gamma p$ candidates, 
 the three-cluster events were analyzed assuming that the final-state
 proton was not detected.
 
 The selection of candidate events and the reconstruction of the reaction
 kinematics were based on the kinematic-fit technique.
 Details of the kinematic-fit parametrization of the detector
 information and resolutions are given in Ref.~\cite{slopemamic}.
 Because e/m showers from electrons and positrons are
 very similar to those of photons, 
 the hypothesis $\gamma p \to 3\gamma p$ was tested to identify
 the $\gamma p\to e^+e^-\gamma p$ candidates.
 The events that satisfied this hypothesis with the probability
 (or confidence level, CL) greater
 than 1\% were accepted for further analysis. The kinematic-fit output
 was used to reconstruct the kinematics of the outgoing particles.
 In this output, there was no separation between e/m showers
 caused by the outgoing photon, electron, or positron.  
 In the further analysis the separation of $e^+e^-$ pairs from
 final-state photons was based on the information from the PID detector.
 This procedure was optimized by using a Monte Carlo (MC) simulation of
 the signal events and the main background reactions.

 To minimize systematic uncertainties in the determination
 of experimental acceptances and to measure the TFF energy dependence properly,
 the MC simulation of the signal events was made to be as close as
 possible to the behavior of real $\gamma p\to \pi^0 p \to e^+e^-\gamma p$ events.
 To reproduce the experimental yield of $\pi^0$ mesons and
 their center-of-mass angular distributions as a function of the incident-photon
 energy, the $\gamma p\to \pi^0 p \to \gamma\gamma p$ reaction was measured in
 the same experiment for every individual Tagger channel by using
 the $\pi^0 \to \gamma\gamma$ decay and then used as an input in the MC event generator.
 The results obtained here for the $\pi^0$ photoproduction angular distributions in
 the center-of-mass frame at the energies of each Tagger channel were
 in good agreement with the previous A2 results for the differential cross sections
 of $\pi^0$ photoproduction on the free proton~\cite{hornidge13,pi0_a2_2015}
 and existing partial-wave analyses (PWA). 
\begin{figure}
\includegraphics[width=0.5\textwidth]{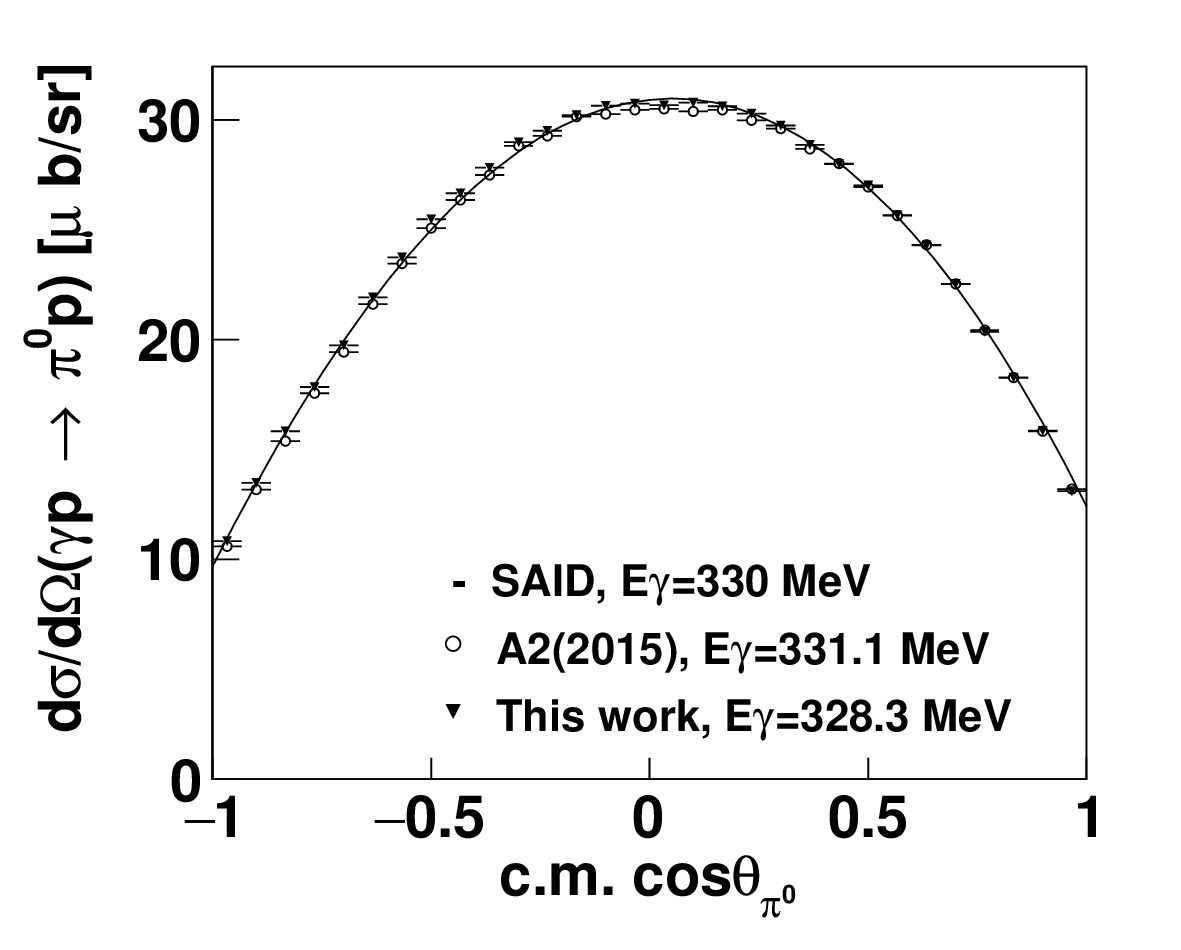}
\caption{
  The $\gamma p\to \pi^0 p$ differential cross sections as a function of the $\pi^0$
  production angle in the center-of-mass frame are shown for the photon-beam energies
  $E_\gamma=328.3$~MeV from the present work (solid triangles),
  $E_\gamma=331.1$~MeV from the most recent A2 measurement~\protect\cite{pi0_a2_2015}
  (open circles), and $E_\gamma=330.0$~MeV from one of the previous PWA
  solutions~\protect\cite{cm12} by SAID~\protect\cite{SAID} (solid line).
  The horizontal error bars represent the bin size for $\cos\theta$. The vertical
  error bars, representing statistical uncertainties are very small and don't
  exceed the size of the data-point markers.
}
 \label{fig:CosThPi0_apr13_2018_eg330} 
\end{figure}
 Such an agreement is illustrated in Fig.~\ref{fig:CosThPi0_apr13_2018_eg330}
 for the energy region corresponding to the largest total cross section near
 the photon-beam energy $E_\gamma=330.0$~MeV. A small difference in $E_\gamma$
 from different experiments is due to different energy of MAMI for its electron beam
 and due to the use of different tagging spectrometers in those experiments.
 As shown, the data points from the two different data sets are practically overlap,
 and a very small difference between the them could be explained by
 their small difference in $E_\gamma$.   

 The $\pi^0 \to e^+e^-\gamma$ decays were generated according to Eq.(\ref{eqn:dgdm_pi0}),
 with the phase-space term removed and with the slope parameter of
 the $\pi^0$ electromagnetic transition form factor from Eq.(\ref{eqn:Fm}) taken as
 $a_\pi = 0.032$, which is close to the value expected from the theoretical
 calculations~\cite{Mas12,Hoferichter_2018,Hoferichter_2018_2}. Such a value slightly increase the population
 of MC events at large $m(e^+e^-)$ masses, which are needed for the acceptance determination,
 but does not affect its value extracted from the analysis of the corresponding MC simulations. 
 The angular dependence of the virtual photon decaying into the $e^+e^-$ pair
 was generated according to Eq.(\ref{eqn:dtheta}).
 These dependences from the leading-order QED term of the decay amplitude
 were then convoluted with radiative corrections based on the calculations
 of Ref.~\cite{Husek_2015}.
 The event vertices were generated uniformly along the 10-cm length of the LH$_2$ target.
 Similarly to the $\gamma p\to \pi^0 p \to e^+e^-\gamma p$ process, the MC generator for
 the main background process $\gamma p\to \pi^0p\to \gamma\gamma p$ used the experimental
 spectra measured in the same experiment.

 For both $\pi^0$ decay modes, the generated events
 were propagated through a \textsc{GEANT} (version 3.21) simulation of the experimental
 setup. To reproduce the resolutions observed in the experimental data, the \textsc{GEANT}
 output (energy and timing) was subject to additional smearing, thus
 allowing both the simulated and experimental data to be analyzed in the same way.
 Matching the energy resolution between the experimental and MC events
 was achieved by adjusting the invariant-mass resolutions,
 the kinematic-fit stretch functions (or pulls), and probability
 distributions. Such an adjustment was based on the analysis of the
 same data sets for the reaction $\gamma p\to \pi^0 p\to \gamma\gamma p$,
 having almost no background from other physical reactions at these energies.
 The simulated events were also tested to check whether they passed
 the trigger requirements.
 
 The PID detector was used to identify the final-state $e^+e^-$ pair
 in the events initially selected as $\gamma p \to 3\gamma p$ candidates.
 Only events with three e/m showers in the CB were selected for further analysis
 becausethe PID provides full coverage of the LH$_2$ target solely for the CB.   
 The identification of $e^{\pm}$ in the CB was based on a correlation
 between azimuthal angles of fired PID elements with the corresponding angles
 of e/m showers in the calorimeter.
 The MC simulations of $\gamma p\to \pi^0 p \to e^+e^-\gamma p$ and
 $\gamma p\to \pi^0p\to \gamma\gamma p$
 were used to optimize this procedure, minimizing the probability for misidentifing
 $e^{\pm}$ with the final-state photons and protons and suppressing the major background
 process.

 The MC simulation for the main background reaction $\gamma p\to \pi^0 p \to \gamma\gamma p$
 demonstrated that this process can mimic $\pi^0 \to e^+e^-\gamma$ events when one of
 the final-state photons is converted into an $e^+e^-$ pair in the material between
 the production vertex and the NaI(Tl) surface. The reconstructed $m(e^+e^-\gamma)$
 distribution for such a background is typically peaked near the $\pi^0$ mass.
 Because the opening angle between the conversion electron and positron
 is typically very small, this background contributes mostly to low invariant
 masses $m(e^+e^-)$. Requiring $e^+$ and $e^-$ to be identified by different PID
 elements suppresses this background significantly. Such a requirement also results
 in losses of actual $\pi^0 \to e^+e^-\gamma$ events. However, it mostly affects
 low invariant masses $m(e^+e^-)$, which are less important for measuring
 the TFF slope parameter. 

 For both the $\gamma p\to \pi^0 p \to e^+e^-\gamma p$ and
 $\gamma p\to \pi^0p\to \gamma\gamma p$ reactions, there is a chance that the recoil proton
 is misidentified as $e^{\pm}$. Such a background does not mimic the $\pi^0$ peak
 in the $m(e^+e^-\gamma)$ spectrum, but its suppression improves the signal-to-background
 ratio, which is important for more reliable fitting of the signal peak above
 the remaining background. The background from the misidentification of the recoil proton
 with $e^{\pm}$ can be suppressed by the analysis of energy losses,
 $dE/dx$, in the PID elements. To reflect the actual differential energy
 deposit $dE/dx$ in the PID, the energy signal from each element,
 ascribed to either $e^+$ or $e^-$, was multiplied by the sine
 of the polar angle of the corresponding particle,
 the magnitude of which was taken from the kinematic-fit output.
 All PID elements were calibrated so that the $e^{\pm}$ peak position matched
 the corresponding peak in the MC simulation.
 To reproduce the actual energy resolution of the PID with the MC simulation,  
 the \textsc{GEANT} output for PID energies was subject to additional smearing,
 allowing the $e^{\pm}$ selection with $dE/dx$ cuts to be very similar for
 the experimental and MC simulated data. The PID energy resolution in the MC simulations
 was adjusted to match the experimental $dE/dx$ spectra for the $e^{\pm}$ particles
 from $\pi^0 \to e^+e^-\gamma$ decays observed experimentally.
 Possible systematic uncertainties due to the $dE/dx$ cuts were checked
 via the stability of the results after narrowing the $dE/dx$ range for selecting $e^{\pm}$.
 Additional information on using the $dE/dx$ PID for event selection and the agreement
 between the experimental data and MC simulations can be found in Ref.~\cite{Pi0_TFF_A2_2016}.
\begin{figure}
\includegraphics[width=0.5\textwidth,height=4.6cm]{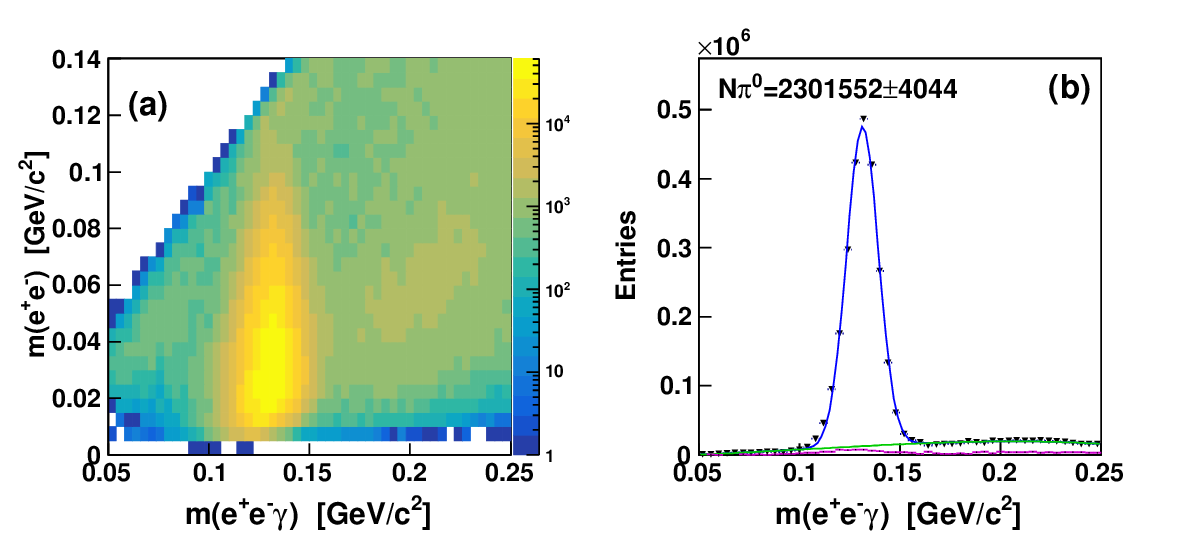}
\caption{ (Color online)
 Experimental events selected as $\gamma p\to e^+e^-\gamma p$ candidates by
 requiring the kinematic-fit CL($\gamma p\to 3\gamma p$)$>$1\% and $e^{\pm}$ to be identified
 by using $dE/dx$ from the PID to suppress their misidentification with recoiling protons:
 (a) the two-dimensional distribution of $m(e^+e^-\gamma)$ vs $m(e^+e^-)$;
 (b) the $m(e^+e^-\gamma)$ distribution fitted with the sum of a Gaussian for
 the $\pi^0 \to e^+e^-\gamma$ peak (blue line) and a polynomial of order 4 for
 the background (green line), and the subtracted empty-target background shown
 by the magenta line.
}
 \label{fig:Eegz34_pi0_2018_2x1} 
\end{figure}
\begin{figure*}
\includegraphics[width=1.\textwidth]{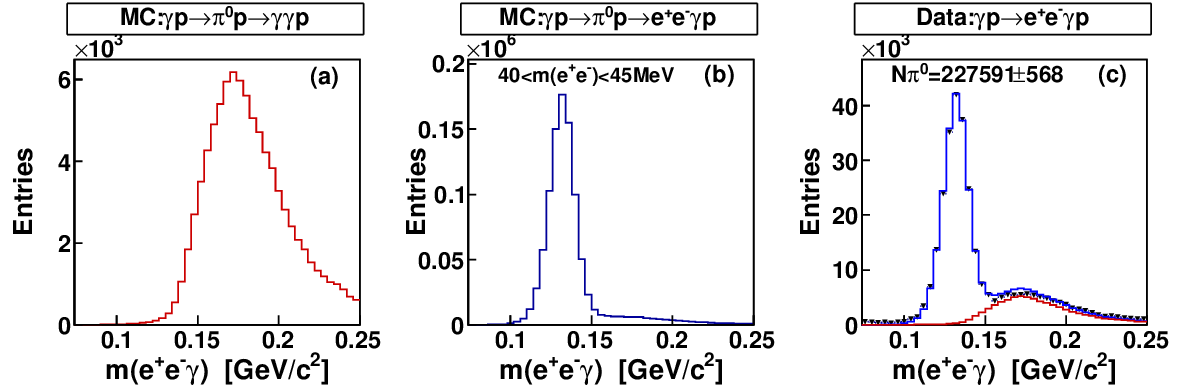}
\caption{ (Color online)
 $m(e^+e^-\gamma)$ invariant-mass distributions obtained in
 the range $40<m(e^+e^-)<45$~MeV/$c^2$ with $\gamma p\to e^+e^-\gamma p$ candidates
 selected without using a $dE/dx$ PID cut to separate $e^{\pm}$ from the recoil protons:
 (a)~MC simulation of $6.2\times10^9$ $\gamma p\to \pi^0 p \to \gamma\gamma p$ events;
 (b)~MC simulation of $1.6\times10^8$ $\gamma p\to \pi^0 p \to e^+e^-\gamma p$ events;
 (c)~experimental spectrum (solid triangles with error bars) fitted with the sum
     of the $\gamma p\to \pi^0 p \to e^+e^-\gamma p$ and $\gamma p\to \pi^0 p \to \gamma\gamma p$
     MC simulations (shown by the blue line), with the fraction of
     the $\gamma p\to \pi^0 p \to \gamma\gamma p$ background shown by the red line. 
}
 \label{fig:pi0eeg_2018_mee40_45mev_hfit_3x1} 
\end{figure*}
\begin{figure*}
\includegraphics[width=1.\textwidth]{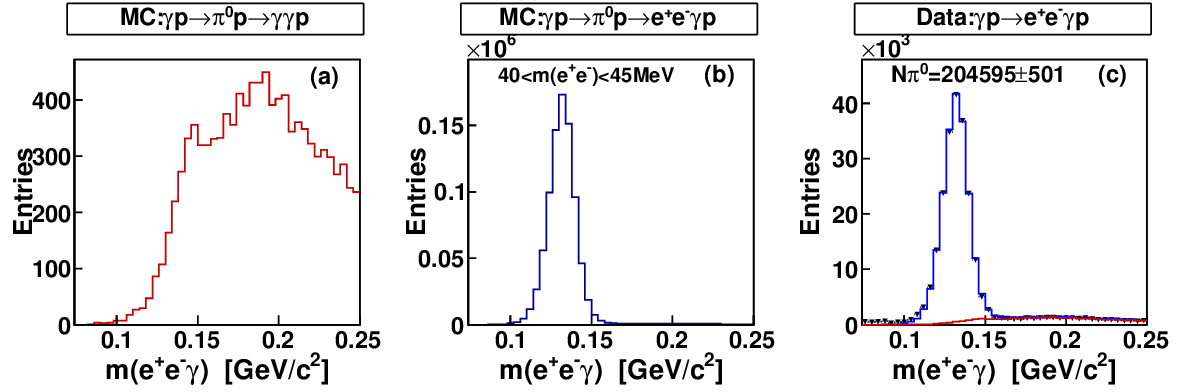}
\caption{ (Color online)
  Same as Fig.~\protect\ref{fig:pi0eeg_2018_mee40_45mev_hfit_3x1}, but with
  the use of a $dE/dx$ PID cut to separate $e^{\pm}$ from the recoil protons.
}
 \label{fig:pi0eeg_2018_mee40_45mev_epidcut_hfit_3x1} 
\end{figure*}

 In addition to the background contributions discussed above,
 there are two more of those that can be subtracted directly in the data analysis.
 The first contribution comes from interactions of incident photons in the windows
 of the target cell. The subtraction of this background was based on the
 analysis of data samples that were taken with an empty target. The weight for
 the subtraction of the empty-target spectra was taken as a ratio of
 the photon-beam fluxes for the data samples with the full and the empty target.
 Another background was caused by random coincidences
 of the Tagger counts with the experimental trigger;
 its subtraction was carried out by using 
 event samples for which all coincidences were random
 (see Refs.~\cite{slopemamic,etamamic} for more details).

\subsection{Analysis of $\pi^0 \to e^+e^-\gamma$ decays}
\label{subsec:Data-II}

 The two-dimensional distribution of $m(e^+e^-\gamma)$ vs $m(e^+e^-)$
 is shown in Fig.~\ref{fig:Eegz34_pi0_2018_2x1}(a) for all selected experimental
 events after subtracting the empty-target and random backgrounds and suppressing
 the misidentification of the recoil protons with $e^{\pm}$ by the $dE/dx$ PID cuts.
 The actual $\pi^0 \to e^+e^-\gamma$ decays are seen there as a vertical band along
 $m(e^+e^-\gamma)$ with its invariant-mass spectrum peaking at the $\pi^0$ mass.
 The corresponding $m(e^+e^-\gamma)$ projection is shown in
 Fig.~\ref{fig:Eegz34_pi0_2018_2x1}(b). To estimate the number of $\pi^0 \to e^+e^-\gamma$
 decays detected in the present experiment, the $m(e^+e^-\gamma)$ distribution
 was fitted with the sum of a Gaussian for the $\pi^0 \to e^+e^-\gamma$ peak
 (blue line) and a polynomial of order 4 for the background (green line).
 The subtracted empty-target background is shown in the same figure by a magenta line.
 Based on this fit, the total number of the detected $\pi^0 \to e^+e^-\gamma$ decays
 is $\approx2.3\times10^6$, which is significantly larger than any of those numbers observed in
 the previous measurements of this decay~\cite{Pi0_TFF_A2_2016,Pi0_TFF_NA62_2017}.
 The background under the $\pi^0 \to e^+e^-\gamma$ peak nearly all comes from
 $\pi^0 \to \gamma\gamma$ decays. Based on the number of $\gamma p\to \pi^0 p \to \gamma\gamma p$
 events observed in this experiment ($3.27\times10^9$), the fraction of those events
 that contributed to this background is only 0.02\%.
 
 To measure the $\pi^0 \to e^+e^-\gamma$ yield as a function of
 the invariant mass $m(e^+e^-)$, the selected
 events were divided into 5-MeV-wide $m(e^+e^-)$ bins.
 Events with $m(e^+e^-)<15$~MeV/$c^2$ were not analyzed
 because e/m showers from those $e^+$ and $e^-$ often overlap in the CB
 and hit the same PID element.
 The number of $\pi^0 \to e^+e^-\gamma$ decays
 in every $m(e^+e^-)$ bin was determined by fitting the $m(e^+e^-\gamma)$ spectra obtained
 from the analysis of the MC simulations for the $\pi^0 \to e^+e^-\gamma$ and
 $\pi^0 \to \gamma\gamma$ decays to the corresponding experimental spectra by using
 the binned maximum-likelihood method.
 Such an approach helps to describe better the signal peak as well as the background
 shape from the events in which the recoil proton was misidentified with $e^{\pm}$.
 To perform the binned maximum-likelihood fits, the {\sc TFractionFitter} class of
 the {\sc Root} CERN software library was used, which were modified
 from the original {\sc HMCMLL} program~\cite{HMCMLL} of the {\sc HBOOK} library, with
 the fraction errors representing $1\sigma$ confidence interval.
 The use of only two MC-simulation fractions to describe the experimental spectra
 resulted in the fit with just one parameter $P_1$ for the signal fraction,
 with the background fraction then defined by $1.-P_1$, in such a way avoiding a correlation
 between different fit parameters.
\begin{figure*}
\includegraphics[width=1.\textwidth]{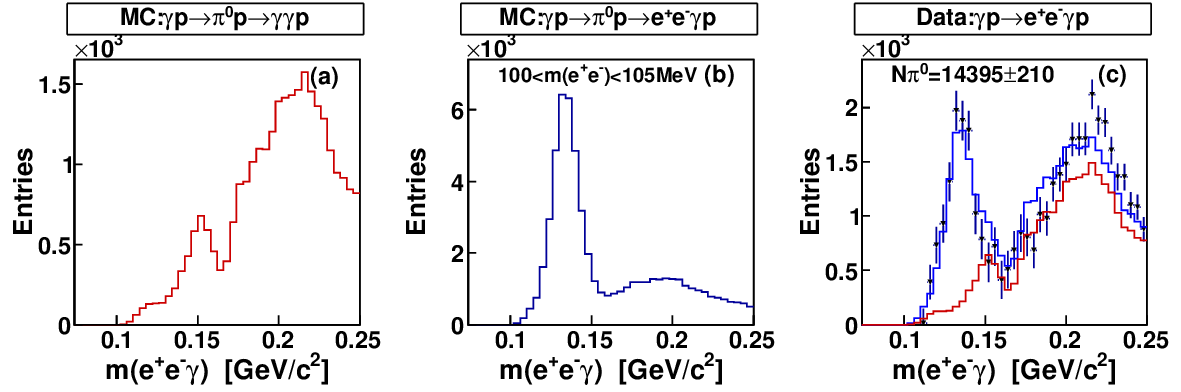}
\caption{ (Color online)
  Same as Fig.~\protect\ref{fig:pi0eeg_2018_mee40_45mev_hfit_3x1}, but
  for the invariant-mass range $100<m(e^+e^-)<105$~MeV/$c^2$.
}
 \label{fig:pi0eeg_2018_mee100_105mev_hfit_3x1} 
\end{figure*}
\begin{figure*}
\includegraphics[width=1.\textwidth]{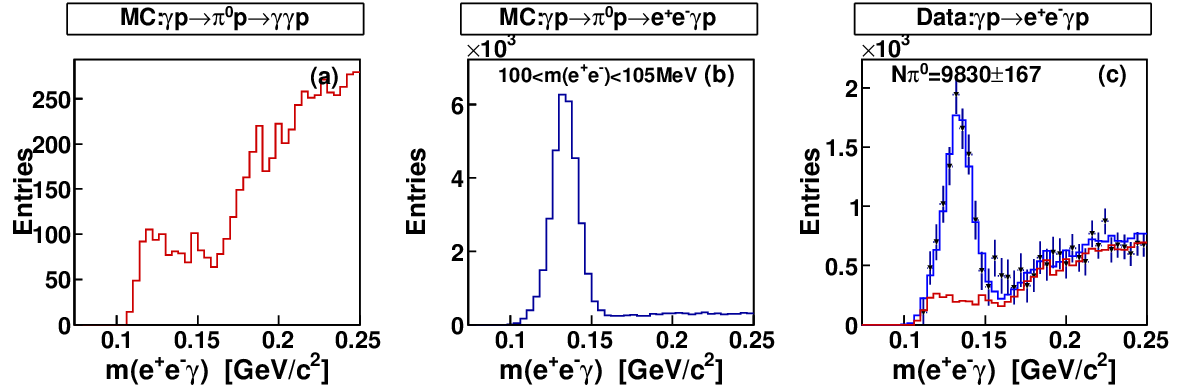}
\caption{ (Color online)
  Same as Fig.~\protect\ref{fig:pi0eeg_2018_mee100_105mev_hfit_3x1}, but with
  using a $dE/dx$ PID cut to separate $e^{\pm}$ from the recoil protons.
}
 \label{fig:pi0eeg_2018_mee100_105mev_epidcut_hfit_3x1} 
\end{figure*}

 The fitting procedure for measuring the number of reconstructed $\pi^0 \to e^+e^-\gamma$ decays
 and the impact of applying a $dE/dx$ PID cut on the fraction of background events
 is illustrated in
 Figs.~\ref{fig:pi0eeg_2018_mee40_45mev_hfit_3x1}--\ref{fig:pi0eeg_2018_mee100_105mev_epidcut_hfit_3x1}
 for two different $m(e^+e^-)$ bins. 
 Figure~\ref{fig:pi0eeg_2018_mee40_45mev_hfit_3x1} shows the $m(e^+e^-\gamma)$ invariant-mass spectra
 obtained for the MC simulations of $\gamma p\to \pi^0 p \to \gamma\gamma p$ and
 $\gamma p\to \pi^0 p \to e^+e^-\gamma p$ and their fit to the experimental distribution reconstructed
 in the range $40<m(e^+e^-)<45$~MeV/$c^2$. There was no $dE/dx$ PID cut applied to obtain these spectra. 
 Figure~\ref{fig:pi0eeg_2018_mee40_45mev_epidcut_hfit_3x1} illustrates the impact of the $dE/dx$ PID cut
 on the fraction of background events in which the recoil proton was misidentified with $e^{\pm}$.
 As shown, applying the $dE/dx$ PID cut reduces the background from
 $\gamma p\to \pi^0 p \to \gamma\gamma p$ by more than one order of magnitude.
 For $\gamma p\to \pi^0 p \to e^+e^-\gamma p$ events, the same cut practically eliminates all
 background under the $\pi^0$ peak, with a quite small reduction of the $\pi^0$ peak itself.
 The fit of the two MC simulations $\gamma p\to \pi^0 p \to \gamma\gamma p$ and
 $\gamma p\to \pi^0 p \to e^+e^-\gamma p$ to the experimental spectrum appears to be sufficient
 for its good description. The agreement observed between the experimental and MC peaks from
 $\pi^0 \to e^+e^-\gamma$ decays confirms the agreement of the experimental data
 and the MC simulations in the energy calibration and resolution of the calorimeters.
 Figures~\ref{fig:pi0eeg_2018_mee100_105mev_hfit_3x1} and \ref{fig:pi0eeg_2018_mee100_105mev_epidcut_hfit_3x1}
 illustrate the fitting procedure of the $m(e^+e^-\gamma)$ invariant-mass spectra
 in the range of high $m(e^+e^-)$ masses,
 where the signal-to-background ratio drops significantly, again obtained before and after applying
 the $dE/dx$ PID cut. As shown, applying the $dE/dx$ PID cut in the range of high $m(e^+e^-)$ masses
 still leaves a significant fraction of background events under the $\pi^0$ peak. 
\begin{figure*}
\includegraphics[width=1.\textwidth]{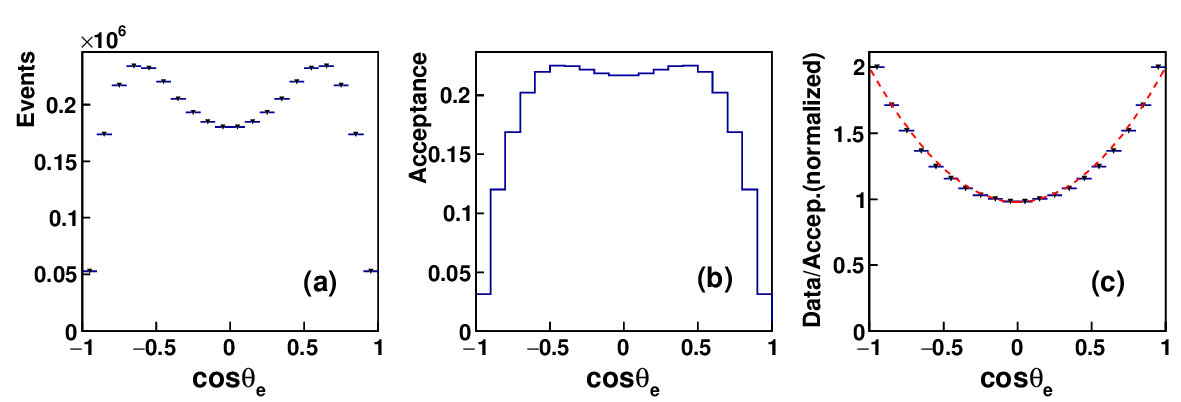}
\caption{ (Color online)
 The $\pi^0\to \gamma\gamma^* \to \gamma e^+e^-$ angular dependence
 (in the $\pi^0$ rest frame)
 of the virtual photon decaying into a $e^+e^-$ pair, with $\theta_e$
 being the angle between the direction of one of the leptons in
 the virtual-photon (or the dilepton) rest frame and the direction
 of the dilepton system (which is opposite to the $\gamma$ direction):
 (a) experimental events from the $\pi^0\to \gamma e^+e^-$ peak;
 (b) angular acceptance based on the MC simulation;
 (c) the experimental spectrum corrected for the acceptance and normalized to
 the corresponding theoretical prediction (shown by a red dashed line).
 Because $e^+$ and $e^-$ cannot be separated in the present experiment,
 the angles of both leptons were used, resulting in a symmetric shape
 with respect to $\cos\theta_e=0.$
}
 \label{fig:eeg_pi0_cth2e_2018} 
\end{figure*}

 The invariant-mass range $15<m(e^+e^-)<60$~MeV/$c^2$, which has the smallest fraction
 of background events under the $\pi^0$ peak, was also used to measure 
 the $\pi^0\to \gamma\gamma^* \to \gamma e^+e^-$ angular dependence $\cos\theta_e$
 of the virtual photon decaying into an $ e^+e^-$ pair and to compare it to the dependence
 determined by Eq.(\ref{eqn:dtheta}) folded with the corresponding radiative
 corrections~\cite{Husek_2015}.
 The experimental results for such an angular dependence are
 illustrated in Fig.~\ref{fig:eeg_pi0_cth2e_2018}, which depicts
 the experimental distribution in
 Fig.~\ref{fig:eeg_pi0_cth2e_2018}(a),
 the angular acceptance determined from the MC simulation
 in Fig.~\ref{fig:eeg_pi0_cth2e_2018}(b), and
 the experimental distribution corrected for the acceptance and compared
 to the theoretical prediction in Fig.~\ref{fig:eeg_pi0_cth2e_2018}(c).
 As shown, in general there is good agreement between the measured and the predicted
 angular dependence. The very small difference between the reconstructed angular distribution
 and its theoretical prediction may be due to a small background contribution still remaining
 in the experimental data and/or due to smearing effects caused by the experimental
 angular resolution. Because $e^+$ and $e^-$ cannot be separated in the present experiment,
 the angles of both leptons were used to measure the angular dependence of the decaying
 dilepton, which resulted in a symmetric shape with respect to $\cos\theta_e=0.$

\section{Results and discussion}
  \label{sec:Results}

\begin{figure}
\includegraphics[width=0.5\textwidth]{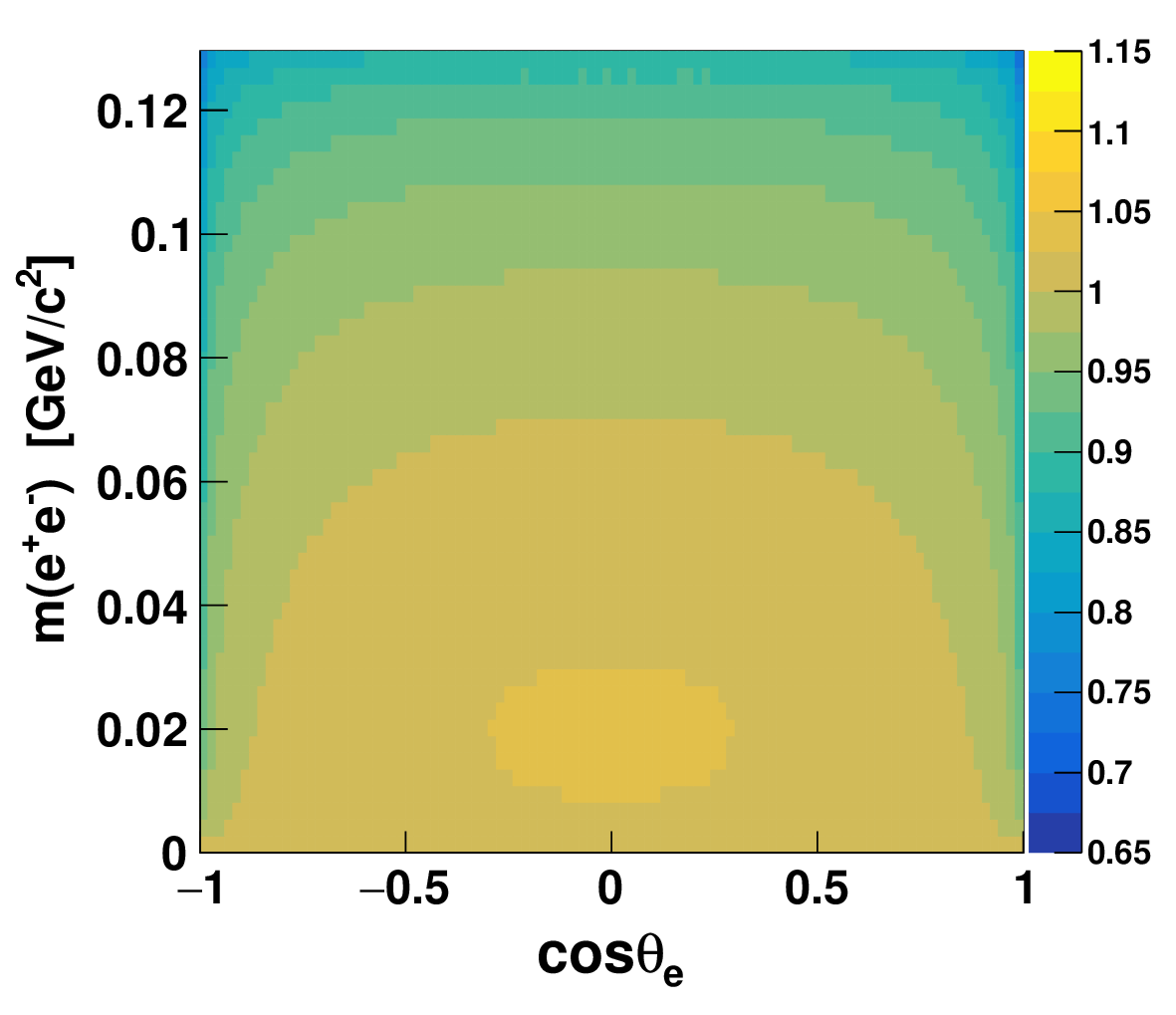}
\caption{ (Color online)
  Magnitude of radiative corrections $\delta(m_{ee},\cos\theta_e)$ from
  Ref.~\protect\cite{Husek_2015} to the $[{\rm{QED}}(m_{ee},\cos\theta_e)]$
  term of Eq.(\protect\ref{eqn:dgdm_pi0_rad}).
}
 \label{fig:pi0eeg_rad_cor} 
\end{figure}

 The total number of $\pi^0 \to e^+e^-\gamma$
 decays produced in each $m(e^+e^-)$ bin was obtained
 by correcting the number of decays observed in those bins
 with the corresponding detection efficiency. The detection efficiency itself
 depends strongly on the selection criteria applied.
 For the initial selection criteria, the detection efficiency rises from 7.06\%
 at $15<m(e^+e^-)<20$~MeV/$c^2$ to 25.9\% at $40<m(e^+e^-)<45$~MeV/$c^2$. 
 The results for $|F_{\pi^0\gamma}(m_{e^+e^-})|^2$ were obtained
 from the numbers of $\pi^0 \to e^+e^-\gamma$ decays produced in each $m(e^+e^-)$ bin,
 taking into account the total number of $\pi^0\to \gamma\gamma$ decays
 produced in the same data
 and the $[{\rm{QED}}(m_{ee})]$ term from Eq.(\ref{eqn:dgdm_pi0_rad})
 after applying radiative corrections according to
 the calculations of Ref.~\cite{Husek_2015}. The magnitude of radiative corrections
 as a two-dimensional function $\delta(m_{ee},\cos\theta_e)$ is depicted
 in Fig.~\ref{fig:pi0eeg_rad_cor}. As shown, radiative corrections make the angular dependence
 of the virtual-photon decay weaker. For the $\pi^0$ Dalitz decay,
 the corrected $[{\rm{QED}}(m_{ee})]$ term integrated over $\cos\theta_e$ becomes $\approx$1\%
 larger at $m(e^+e^-)=15$~MeV, and $\approx$10\% lower at $m(e^+e^-)=120$~MeV.
 Such radiative corrections make the $|F_{\pi^0\gamma}(m_{e^+e^-})|^2$ results at high $m(e^+e^-)$
 masses very sensitive to the magnitude of those corrections and to a possible impact on it
 from next-to-next-to-leading-order (NNLO) terms, which were not taken into account in the
 calculations of Ref.~\cite{Husek_2015}. It was also checked that the $|F_{\pi^0\gamma}(m_{e^+e^-})|^2$
 results do not follow directly the change in the magnitude of radiative corrections,
 because the experimental acceptance slightly improves when those corrections are applied,
 especially at high $m(e^+e^-)$ masses. Such an improvement is because fewer events are remaining
 at the extreme $\theta_e$, where the experimental acceptance drops
 (see Fig.~\ref{fig:eeg_pi0_cth2e_2018}(b)). 
 
 The uncertainty in an individual $|F_{\pi^0\gamma}(m_{e^+e^-})|^2$
 value from a particular fit was based on the uncertainty in the number
 of $\pi^0 \to e^+e^-\gamma$ decays determined by this fit. As it was
 described above, the binned maximum-likelihood method was used to fit
 the $m(e^+e^-\gamma)$ spectra obtained
 from the analysis of the MC simulations for the $\pi^0 \to e^+e^-\gamma$ and
 $\pi^0 \to \gamma\gamma$ decays to the corresponding experimental spectra, where
 the fraction errors from the fit represented $1\sigma$ confidence interval.
 With using parameter $P_1$ to describe the signal fraction, the background
 fraction was described by $1.-P_1$. The uncertainty $\delta N(\pi^0\to e^+e^-\gamma)$ in the number
 of signal events from an individual fit was based then on the fit error for parameter $P_1$.
 For the most part, its magnitude is determined by the experimental statistics for the fraction of
 signal events, but the larger background fraction makes this uncertainty larger as well.
 A similarity in the shape of the signal and background spectra also increases this uncertainty.
 For simplicity, the $|F_{\pi^0\gamma}(m_{e^+e^-})|^2$ uncertainties obtained from individual
 fits are called statistical here.

 For each individual $m(e^+e^-)$ bin, the fitting procedure was repeated multiple
 times after refilling the $m(e^+e^-\gamma)$ spectra with different
 combinations of selection criteria, which were used to improve
 the signal-to-background ratio. The changes in selection criteria
 included different cuts on PID $dE/dx$, the kinematic-fit CL
 (such as 1\%, 2\%, 5\%, and 10\%), the incident-photon energy range
 (excluding higher energies with larger cross sections for the reactions
 $\gamma p\to \pi^0\pi^0 p$ and $\gamma p\to \pi^0\pi^+ n$). 
 For a particular $m(e^+e^-)$ bin, the fit result with the smallest
 $\delta N(\pi^0\to e^+e^-\gamma)/N(\pi^0\to e^+e^-\gamma)$ was taken as
 the main result, in order to provide the smallest uncertainty in the slope parameter
 from fitting to the $|F_{\pi^0\gamma}(m_{e^+e^-})|^2$ data points.
 The other results were used to evaluate the systematic
 uncertainty in the measured value for the slope parameter of the $\pi^0$ e/m TFF.
 Typically the smallest $\delta N(\pi^0\to e^+e^-\gamma)/N(\pi^0\to e^+e^-\gamma)$ were
 obtained for those spectra that had the largest number of the signal events, which was
 associated with looser selection cuts for background events. Such an approach helps
 to avoid introducing possible systematic effects due to tighter selection criteria.

\begin{figure*}
\includegraphics[width=1.\textwidth]{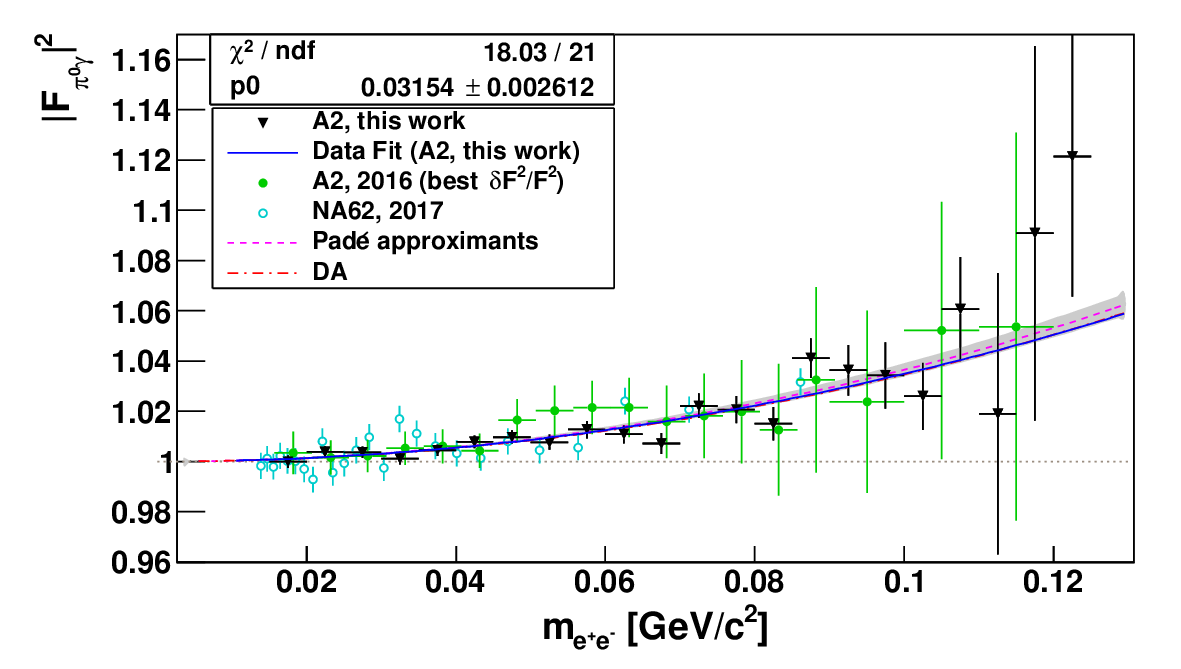}
\caption{ (Color online)
 $|F_{\pi^0\gamma}|^2$ results (black filled triangles) obtained in the present
  work are compared to the data points of the previous measurements
  by the A2~\protect\cite{Pi0_TFF_A2_2016} (green filled circles)
  and NA62~\protect\cite{Pi0_TFF_NA62_2017} (cyan circles) Collaborations
  and to the calculations with Pad\'e approximants~\protect\cite{Mas12}
 (shown by the short-dashed magenta line with an error band)
 and to the dispersive analysis (DA) from Ref.~\protect\cite{Hoferichter_2018,Hoferichter_2018_2}
 (long-dashed red line). The DA error band is not shown here because
 of its smallness. The vertical error bars on all data points are based on the errors
 obtained in the fits to the $m(e^+e^-\gamma)$ experimental spectra and are mostly
 reflect the experimental statistics. The horizontal error bars of the A2 data points
 show the width of the $m(e^+e^-)$ bins. The fit of Eq.(\protect\ref{eqn:Fm}) (by using
 the standard {\sc ROOT}-based {\sc MIGRAD} program)
 to the present $|F_{\pi^0\gamma}|^2$ data points is shown by the blue line.
 The fit parameter $p_0$ represents the slope parameter $a_\pi$, and the fit error for $p_0$
 represents the statistical uncertainty of $a_\pi$.
}
 \label{fig:tff_pi0_eeg_a2_2018_na62} 
\end{figure*}
\begin{table*}
\caption 
[tab:pi0tff]{
 Results of this work for the $\pi^0$ TFF, $|F_{\pi^0\gamma}|^2$, as a function of
 the invariant mass $m(e^+e^-)$ obtained for 22 5-MeV-wide $m(e^+e^-)$ bins in
 the invariant-mass range $15<m(e^+e^-)<125$~MeV/$c^2$, where the uncertainties
 based on the fits to the experimental spectra are mostly statistical,
 reflecting the number of $\pi^0 \to e^+e^-\gamma$ decays observed in each bin.
 The number of significant digits in the listed $|F_{\pi^0\gamma}|^2$ value
 was left identical to those used in their fit with the slope parameter.
 The $|F_{\pi^0\gamma}|^2$~(LO) results are obtained with radiative corrections including
 only the LO terms.
} \label{tab:pi0tff}
\begin{ruledtabular}
\begin{tabular}{|c|c|c|c|c|c|} 
\hline
 $m(e^+e^-)$~[MeV/$c^2$]
 & $17.5\pm2.5$ & $22.5\pm2.5$ & $27.5\pm2.5$ & $32.5\pm2.5$ & $37.5\pm2.5$ \\
\hline
  $|F_{\pi^0\gamma}|^2$
 & $0.99998\pm0.00246$ & $1.00382\pm0.00218$ & $1.00375\pm0.00217$ & $1.00110\pm0.00208$ & $1.00450\pm0.00236$ \\
\hline
  $|F_{\pi^0\gamma}|^2$~(LO)
 & $0.99989\pm0.00246$ & $1.00358\pm0.00218$ & $1.00335\pm0.00217$ & $1.00049\pm0.00208$ & $1.00380\pm0.00236$ \\
\hline
\hline
 $m(e^+e^-)$~[MeV/$c^2$]
 & $42.5\pm2.5$ & $47.5\pm2.5$ & $52.5\pm2.5$ & $57.5\pm2.5$ & $62.5\pm2.5$ \\
\hline
  $|F_{\pi^0\gamma}|^2$
 & $1.00779\pm0.00252$ & $1.00973\pm0.00262$ & $1.00772\pm0.00303$ & $1.01272\pm0.00342$ & $1.01086\pm0.00367$ \\
\hline
  $|F_{\pi^0\gamma}|^2$~(LO)
 & $1.00686\pm0.00252$ & $1.00848\pm0.00262$ & $1.00617\pm0.00303$ & $1.01079\pm0.00341$ & $1.00853\pm0.00366$ \\
\hline
\hline
 $m(e^+e^-)$~[MeV/$c^2$]
  & $67.5\pm2.5$ & $72.5\pm2.5$ & $77.5\pm2.5$ & $82.5\pm2.5$ & $87.5\pm2.5$ \\
\hline
  $|F_{\pi^0\gamma}|^2$
& $1.00721\pm0.00421$ & $1.02222\pm0.00486$ & $1.02069\pm0.00543$ & $1.01511\pm0.00659$ & $1.04125\pm0.00785$ \\
\hline
  $|F_{\pi^0\gamma}|^2$~(LO)
& $1.00445\pm0.00420$ & $1.01889\pm0.00484$ & $1.01685\pm0.00541$ & $1.01065\pm0.00656$ & $1.03613\pm0.00781$ \\
\hline
\hline
 $m(e^+e^-)$~[MeV/$c^2$]
  & $92.5\pm2.5$ & $97.5\pm2.5$ & $102.5\pm2.5$ & $107.5\pm2.5$ & $112.5\pm2.5$  \\
\hline
  $|F_{\pi^0\gamma}|^2$
& $1.03634\pm0.01011$ & $1.03428\pm0.01322$ & $1.02597\pm0.01327$ & $1.06060\pm0.02074$ & $1.01904\pm0.05599$ \\
\hline
  $|F_{\pi^0\gamma}|^2$~(LO)
& $1.03064\pm0.01005$ & $1.02967\pm0.01316$ & $1.02050\pm0.01320$ & $1.05562\pm0.02064$ & $1.00732\pm0.05535$ \\
\hline
\hline
 $m(e^+e^-)$~[MeV/$c^2$]
  & $117.5\pm2.5$ & $122.5\pm2.5$ &  &  &  \\
\hline
  $|F_{\pi^0\gamma}|^2$
& $1.09086\pm0.07455$ & $1.12150\pm0.05574$ &  &  & \\
\hline
  $|F_{\pi^0\gamma}|^2$~(LO)
& $1.08401\pm0.07408$ & $1.11955\pm0.05564$ &  &  & \\
\hline
\end{tabular}
\end{ruledtabular}
\end{table*}

 In the previous measurement~\cite{Pi0_TFF_A2_2016} by the A2 Collaboration,
 the main result for a particular $m(e^+e^-)$ bin was obtained by averaging
 the results of all fits made for this bin, in a way that reduced the scattering of
 the measured $|F_{\pi^0\gamma}(m_{e^+e^-})|^2$ values from one $m(e^+e^-)$ bin to another. 
 Then the uncertainty based on the smallest $\delta N(\pi^0\to e^+e^-\gamma)/N(\pi^0\to e^+e^-\gamma)$
 was added in quadrature with the systematic uncertainty calculated as the root mean square
 of the results from all fits made for this bin. However, such an approach turned out to be
 too conservative in the evaluation of the total uncertainties in
 the $|F_{\pi^0\gamma}(m_{e^+e^-})|^2$ values for each individual $m(e^+e^-)$ bin, as
 the magnitude of those total uncertainties looked significantly larger compared
 to the actual scattering of the data points. Moreover, the slope parameter
 obtained from fitting those $|F_{\pi^0\gamma}(m_{e^+e^-})|^2$ data points did not provide
 a separation of the statistical and the systematic components in the total uncertainty
 for the slope parameter.
 
 The $|F_{\pi^0\gamma}(m_{e^+e^-})|^2$ data points from the present work are depicted
 in Fig.~\ref{fig:tff_pi0_eeg_a2_2018_na62} along with the results of previous
 measurements by the A2~\cite{Pi0_TFF_A2_2016} and NA62~\cite{Pi0_TFF_NA62_2017}
 Collaborations and the calculations with Pad\'e approximants~\cite{Mas12} and
 the dispersive analysis (DA) from Ref.~\cite{Hoferichter_2018,Hoferichter_2018_2}.
 The vertical error bars plotted on all data points are based on the errors obtained
 in the fits to the $m(e^+e^-\gamma)$ experimental spectra, mostly reflecting
 the experimental statistics. The horizontal error bars of the A2 data points
 show the width of the $m(e^+e^-)$ bins. In order to facilitate a comparison with the present work,
 the data points shown for the previous A2 measurement~\cite{Pi0_TFF_A2_2016} represent
 the fit results with the smallest $\delta N(\pi^0\to e^+e^-\gamma)/N(\pi^0\to e^+e^-\gamma)$,
 rather than the results averaged for each bin, with the systematic uncertainties added.
 In this way, a better view of the improvement reached by the A2 Collaboration,
 due to the increase in the experimental statistics, is presented.
 The vertical error bars of the NA62 data points
 are of the same order of magnitude because their $m(e^+e^-)$ bins were chosen in such
 a way to have similar experimental statistics in each bin. The magnitude of their error bars
 are typically larger than those from the present work over most of the entire
 range of $m(e^+e^-)$ masses. The fit of Eq.(\protect\ref{eqn:Fm}), by using
 the standard {\sc ROOT}-based {\sc MIGRAD} program,
 to the present $|F_{\pi^0\gamma}|^2$ data points is shown by the blue line.
 The fit parameter $p_0$ represents the slope parameter $a_\pi$, and the fit error for $p_0$
 represents the statistical uncertainty of $a_\pi$.
 Rounding the fit output for parameter $p_0$ to the proper number of significant digits
 results in the value $a_\pi= 0.0315\pm 0.0026_{\mathrm{stat}}$, with its statistical uncertainty
 being close to 8\%. The magnitude of $\chi^2/$ndf=18.03/21 from the fit indicates
 that the scattering of the experimental data points with respect to the fit function
 is in close agreement with their uncertainties.

\begin{figure}
\includegraphics[width=0.5\textwidth]{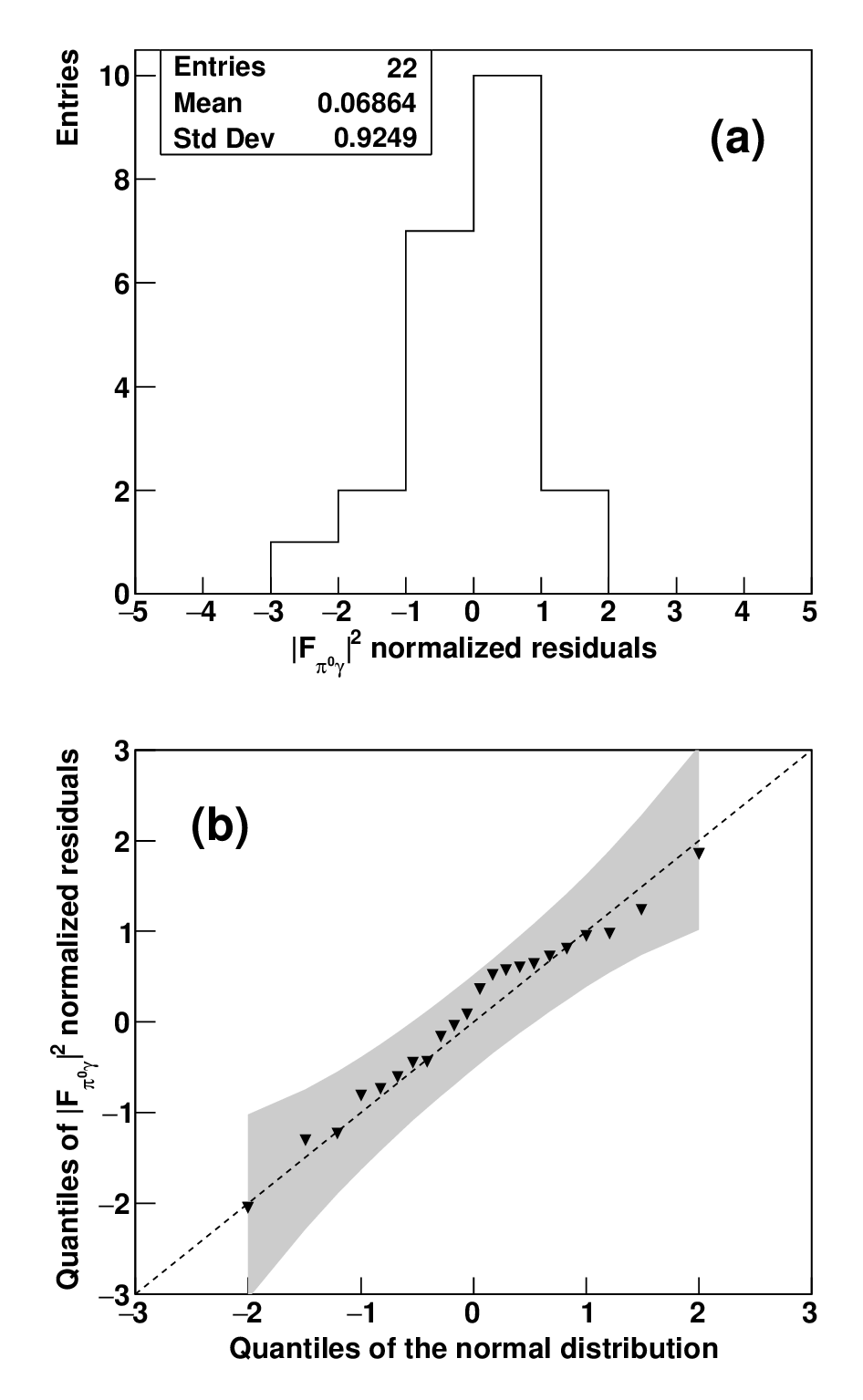}
\caption{
  (a) distribution of the normalized residuals obtained from the fit of
  the slope parameter to the present 22 experimental data points.
  (b) quantiles of the present normalized residuals compared to quantiles for the normal
  distribution of 22 points, with the diagonal 
  dashed line showing the expected dependence if the residuals follow the normal distribution.
  The gray band shows the 95\% ($1.96\sigma$) statistical-uncertainty interval for quantiles of
  the normalized residuals.
}
 \label{fig:tff_pi0_eeg_a2_2018_res_1x2} 
\end{figure}

 An additional analysis of the quality of the fit to the present data points and
 their statistical uncertainties was made by using their normalized residuals.
 For an individual data point, its normalized residual was defined as
 $R(x) = \frac{Y(x) - \widehat{Y}(x)}{\delta Y(x)}$, where $x=m_{e^+e^-}$,
 $Y(x)=|F_{\pi^0\gamma}(m_{e^+e^-})|^2$ is a measured data point with its statistical
 uncertainty $\delta Y(x)$, and $\widehat{Y}(x)$ is an expected value taken from
 the fit to the experimental data points. Then an individual normalized residual
 reflects the distance from the fit to the given data point in the units of standard deviations.
 The own uncertainty of a normalized residual would be 1.0 in the case when
 $\widehat{Y}(x)$ has zero uncertainty.
 Also, if the uncertainties of experimental data points are evaluated correctly,
 the normalized fit residuals should follow the normal distribution.
 The distribution of the normalized residuals for the present 22 data points,
 shown in Fig.~\ref{fig:tff_pi0_eeg_a2_2018_res_1x2}(a), has the mean value 0.06864
 and $\sigma=0.9249$, the parameters of which are close to a normal distribution with
 $\sigma=1.0$ and the mean value equal zero.
 By using Student's t-statistics, the 68\% statistical-uncertainty intervals
 for such a set of normalized residuals were found to be 
 $[-0.14; 0.27]$ and $[0.83; 1.13]$, for the mean value and $\sigma$ respectively,
 indicating that the present normalized residuals are in good agreement with the normal
 distribution.

 Figure~\ref{fig:tff_pi0_eeg_a2_2018_res_1x2}(b) shows
 a $Q-Q$ (Quantile-Quantile) plot that compares the cumulative distribution of the
 present 22 normalized residuals with the corresponding theoretical distribution
 for a normal function. The gray band depicts the 95\% ($1.96\sigma$) uncertainty region
 due to the finite data-set dimension obtained using a Monte-Carlo bootstrap simulation.
 As shown, the obtained $Q-Q$ distribution is in agreement with the expected linear dependence
 (depicted by a dashed line) within the statistical error band.
 Such an agreement provides a clear indication of a good quality of both
 the experimental data points with their uncertainties and their fitting procedure used
 to determine the slope parameter.

 It was also checked that the inclusion of an additional normalization parameter in the fit
 function gives a value $0.9999\pm0.0006$ close to unity, proving that it could be neglected
 for extracting the slope parameter $a_\pi$, to avoid enlarging the $a_\pi$ uncertainty
 because of the correlation between the two parameters. As also shown
 in Fig.~\ref{fig:tff_pi0_eeg_a2_2018_na62}, the slope-parameter fit to the present data
 points agrees very closely with the latest DA calculation~\cite{Hoferichter_2018,Hoferichter_2018_2}
 and is also within the error band of the calculation with Pad\'e approximants~\cite{Mas12}. 

 As discussed above, the fitting procedure for each individual $m(e^+e^-)$ bin
 was repeated multiple times after refilling the $m(e^+e^-\gamma)$ spectra with
 different combinations of selection criteria, which improved
 the signal-to-background ratio, but lost some fraction of the signal events.
 The main tests were performed for events selected with and without PID $dE/dx$ cuts,
 with tightening PID $dE/dx$ cuts and the kinematic-fit CL, and removing events
 with higher incident-photon energies to reduce possible background from the reactions
 $\gamma p\to \pi^0\pi^0 p$ and $\gamma p\to \pi^0\pi^+ n$.
 The selection of those $|F_{\pi^0\gamma}(m_{e^+e^-})|^2$ results for fitting with Eq.(\ref{eqn:Fm})
 was then based on using similar sets of selection criteria for each test to
 see changes in the magnitude obtained for the slope parameter.
 The root mean square of all the test values obtained for $a_\pi$ was taken as its
 systematic uncertainty, resulting in the final value of this work
\begin{equation}
 a_\pi = 0.0315\pm 0.0026_{\mathrm{stat}}\pm 0.0010_{\mathrm{syst}}.
\label{eqn:api_this_work}
\end{equation}
 This result is in good agreement within the experimental uncertainties with the most
 recent measurements by the A2 Collaboration,
 $a_\pi = 0.030\pm 0.010_{\mathrm{tot}}$, and 
 by the NA62 Collaboration,
 $a_\pi = 0.0368\pm 0.0051_{\mathrm{stat}}\pm 0.0025_{\mathrm{syst}} = 0.0368\pm 0.0057_{\mathrm{tot}}$.
 A similar agreement is observed with the calculations from Ref.~\cite{Mas12},
 $a_\pi = 0.0324\pm0.0012_{\mathrm{stat}}\pm0.0019_{\mathrm{syst}}$, 
 and Ref.~\cite{Hoferichter_2018,Hoferichter_2018_2}, $a_\pi = 0.0315\pm0.0009$;
 although the uncertainty obtained for $a_\pi$ in the present measurement
 is still larger than in those calculations.
 At the same time, the uncertainty in the current RPP value,
 $a_\pi = 0.0332\pm0.0029$~\cite{PDG_2024}, is slightly larger than the one
 in the present measurement.
 In addition, the statistical accuracy of the present $|F_{\pi^0\gamma}(m_{e^+e^-})|^2$ data points
 is still insufficient to fit them with Eq.(\ref{eqn:Fm2}), resulting in
 a strong correlation between the slope parameter $a_\pi$ and the curvature parameter $b_\pi$.

 The numerical values obtained for the individual $|F_{\pi^0\gamma}(m_{e^+e^-})|^2$ results in
 each $m(e^+e^-)$ bin are listed in Table~\ref{tab:pi0tff} in order to facilitate their comparison
 with existing data and theoretical calculations and for using in model-independent fits.
 The sensitivity of the present $|F_{\pi^0\gamma}(m_{e^+e^-})|^2$ results and
 the measured value for the slope parameter to the magnitude of radiative
 corrections was tested by removing their three NLO terms, assuming that the
 LO term is well known and that introducing additionally NNLO
 terms could change the results of this work. The results of this test were listed
 in Table~\ref{tab:pi0tff} as $|F_{\pi^0\gamma}(m_{e^+e^-})|^2$~(LO). Because the change in
 the experimental acceptance after removing only NLO terms is quite small,
 the $|F_{\pi^0\gamma}(m_{e^+e^-})|^2$~(LO) data points could be used in case there will be an update
 of radiative corrections for their NLO and NNLO terms.
 Fitting the $|F_{\pi^0\gamma}(m_{e^+e^-})|^2$~(LO) data points with Eq.(\ref{eqn:Fm}) results
 in $a_\pi(\rm{LO}) = 0.0262\pm0.0026$, which is significantly lower than the actual $a_\pi$ obtained
 in this work, confirming a strong sensitivity of the results obtained in this work
 to the magnitude of radiative corrections.

\section{Summary and conclusions}
\label{sec:Conclusion}
 The Dalitz decay $\pi^0 \to e^+e^-\gamma$ has been measured with
 the highest up-to-date statistical accuracy in the $\gamma p\to \pi^0 p$ reaction
 with the A2 tagged-photon facility at the Mainz Microtron, MAMI.
 The value obtained for the slope parameter of the  $\pi^0$ e/m TFF,
 $a_\pi=0.0315\pm 0.0026_{\mathrm{stat}}\pm 0.0010_{\mathrm{syst}}$,
 agrees within the uncertainties with the existing measurements
 of this decay and with the recent theoretical calculations.
 The uncertainty obtained in the value of $a_\pi$ is lower than in previous
 results based on the $\pi^0\to e^+e^-\gamma$ decay.
 The results of this work also include $|F_{\pi^0\gamma}(m_{ee})|^2$ data points
 and their uncertainties, allowing comparison of the
 present data with previous measurements and theoretical calculations or using
 these data in model-independent fits. It is expected that adding the present 
 value for $a_\pi$ in its RPP average will result in further decrease of
 the uncertainty in the RPP average value. However, it will still be insufficient
 for any significant constraining the calculations used to estimate $(g-2)_\mu$ contributions
 from the pion-pole term $a_{\mu}^{\pi^0}$ to the HLbL scattering and the $\pi^0\gamma$
 term $a_{\mu}^{\pi^0\gamma}$ to the HVP correction, which are mostly determined by the data-driven
 calculations of the $\pi^0$ electromagnetic transition form
 factor $a_\pi$~\cite{Hoferichter_2018,Hoferichter_2018_2}, having significantly smaller
 uncertainties.

\section*{Acknowledgments}

 The authors wish to acknowledge the excellent support of the accelerator group and
 operators of MAMI.
 We would like to stress the importance of involvement of Tom\'a\v{s} Husek
 in the present and previous A2 analysis~\cite{Pi0_TFF_A2_2016} in providing a program to calculate
 radiative corrections and helpful discussion of the possible impact of those corrections
 on the magnitude of the slope parameter extracted from the experimental data.
 We would like to thank Bastian Kubis and Pere Masjuan
 for useful discussions and continuous interest in the paper.
 This work was supported by the Deutsche Forschungsgemeinschaft (DFG,
 German Research Foundation) within the Research Unit ``Photon-photon interactions
 in the Standard Model and beyond'' (Project No.458854507 - FOR 5327),
 the European Community-Research
 Infrastructure Activity under the FP6 ``Structuring the European Research Area''
 program (Hadron Physics, Contract No. RII3-CT-2004-506078), Schweizerischer
 Nationalfonds (Contract Nos. 200020-156983, 132799, 121781, 117601, 113511),
 the U.K. Science and Technology Facilities Council (STFC 57071/1, 50727/1, ST/Y000285/1),
the U.S. Department of Energy (Offices of Science and Nuclear Physics,
 Award Nos. DE-FG02-99-ER41110, DE-FG02-88ER40415, DE-FG02-01-ER41194)
 and National Science Foundation (Grant Nos. PHY-1039130, PHY-1714833,PHY-2012940,
 PHY-2310026, IIA-1358175), NSERC of Canada (Grant Nos. 371543-2012, SAPPJ-2015-00023),
 and INFN (Italy). We thank the undergraduate students of Mount Allison University
 and The George Washington University for their assistance.
 
\section*{Data availability}

All data points from the present work have been included in Table~\ref{tab:pi0tff}.

\end{document}